\documentclass[12pt,letterpaper,conference]{IEEEtran}
\onecolumn
\usepackage{amssymb,amsmath}

\usepackage{cite}

 \usepackage{amsbsy}

\ifCLASSINFOpdf
\else
\fi

\usepackage{graphicx}

\usepackage{url}

\begin{document}
%
\title{Exploratory Data Analysis of The Kelvin–Helmholtz instability in Jets}

\author{\IEEEauthorblockN{Santosh Tirunagari}
\IEEEauthorblockA{Department of Computing \& CVSSP\\
University of Surrey\\
Guildford, Surrey GU2 7XH\\
Email: s.tirunagari@surrey.ac.uk}}
\maketitle

\begin{abstract}
The Kelvin–Helmholtz (KH) instability is a fundamental wave instability that is frequently observed
in all kinds of shear layer (jets, wakes, atmospheric air currents etc). The study of KH-instability, coherent flow structures has a major impact in understanding the fundamentals of fluid dynamics. Therefore there is a need for methods that can identify and analyse these structures. In this Final assignment, we use machine-learning methods such as Proper Orthogonal Decomposition (POD) and Dynamic Mode Decomposition (DMD) to analyse the coherent flow structures. We used a 2D co-axial jet as our data, with Reynolds number corresponding to Re: 10,000. Results for POD modes and DMD modes are discussed and compared.

\end{abstract}


%
\IEEEpeerreviewmaketitle

\section{Introduction}
The Kelvin–Helmholtz (KH) instability can occur when there is passive scalar shear in a single continuous fluid, or where there is a passive scalar difference across the interface between two fluids~\cite{miles1959generation}. To gain deeper understanding on  KH coherent flow strcutures, machine-learning methods can be utilised. This field allows computers to adopt behaviors based on training data. These methods recognize complex patterns and makes intelligent decisions based on data~\cite{bishop2006pattern}. These techniques include: reduced order models or dimensionality reduction methods~\cite{paukkeri2011effect}, statistical, and machine vision methods. Dimensionality reduction is a technique used to find a reduced order model on a given data. Such a technique includes feature selection and feature extraction methods. Feature selection is based on selecting a subset of variables which best define the data, whereas feature extraction transforms the data from high-dimensional space to a space of fewer dimensions~\cite{bishop2006pattern}. The data transformation may be linear, as in Proper Orthogonal Decomposition (POD)~\cite{springerlink:10.1007/s00348-007-0347-6}~\cite{vuorinen2013large}~\cite{tirunagari2012analysis} and may be non-linear such as in Dynamic Mode Decomposition (DMD)~\cite{schmid2009dynamic} ~\cite{schmid2010dynamic}~\cite{tirunagari2012analysis}.

\begin{figure}[htp!]
\centering
\begin{tabular}{cc}
\includegraphics[scale=0.5]{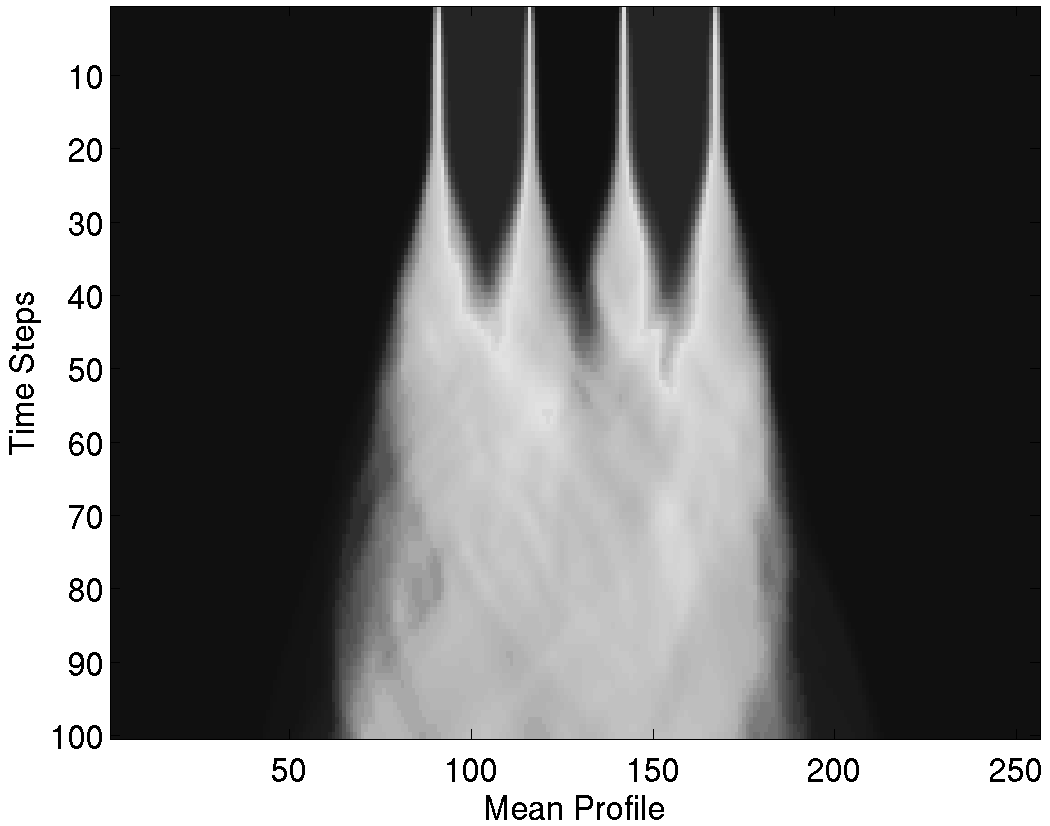} & \includegraphics[scale=0.5]{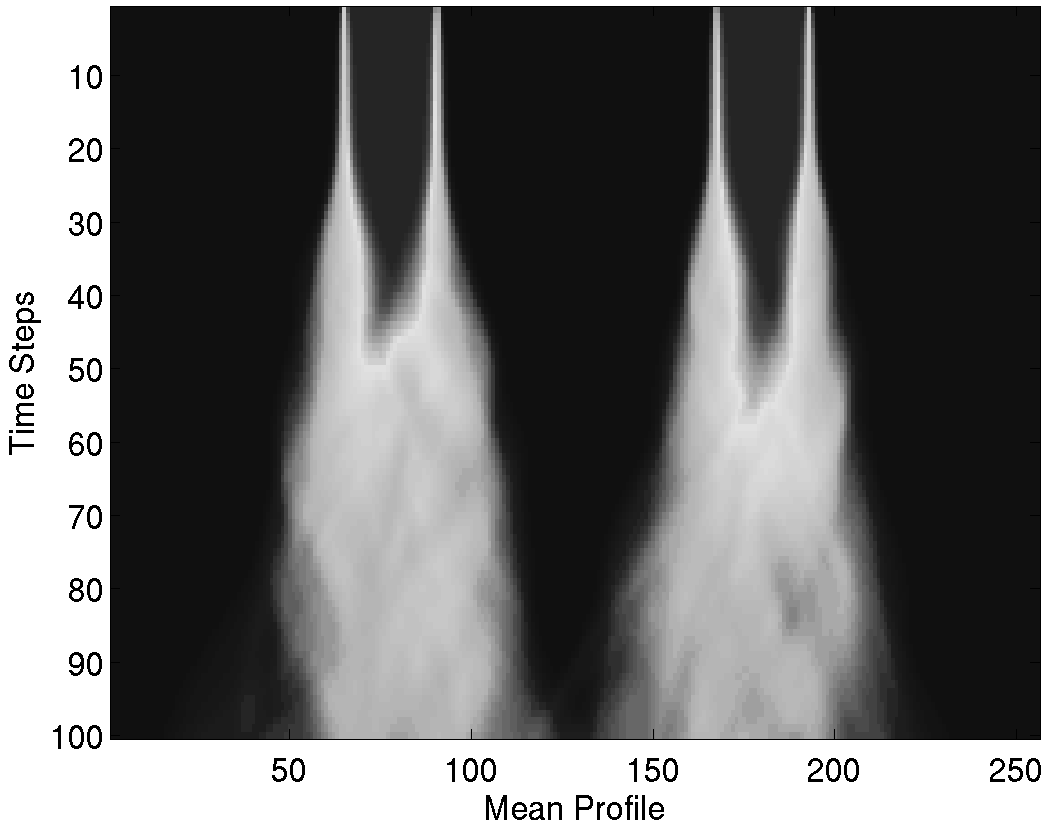}\\
\end{tabular}
\caption{Mean Profiles of  case 1 (left) and case 2 ( right) passive scalar fields}
\label{comples5}
\end{figure}

From the literature, it is expected that POD and DMD would be good methods for analyzing the flow structures. For example, Perrin et. al~\cite{springerlink:10.1007/s00348-007-0347-6} used POD to obtain phase averaged turbulence properties for flow past a cylinder. In highly turbulent flows, the coherent flow structures are difficult to identify due to the combination of organized and chaotic fluctuating motions. Using POD analysis it is shown in~\cite{gorder}, that  von Karman vortices can be reproduced within the first few modes. POD has been used as a tool for the comparison  of Particle Image Velocimetry\footnote{\url{http://en.wikipedia.org/wiki/Particle_image_velocimetry}}(PIV) and Light Eddy Simulation(LES) data in~\cite{meyerandK.E} and it is also shown that POD modes have a good qualitative agreement between PIV and LES. A paper by Schmidt et. al~\cite{schmid2011applications}, used DMD to a sequence of flow images of a slow jet entering quiescent fluid showcased the detection of dynamically relevant coherent structures that play an important role in characterizing the fluid behaviour over processed time interval. To identify the KH coherent flow structures,  we study 2 cases of co-axial jet:
\begin{itemize}
 \item When both the jets are close to each other (CASE = $\frac{L}{10}$).  
 \item When both the jets are independent, far a part from each other (CASE = $\frac{L}{5}$).
\end{itemize}
where $L$ is the domain side length i.e $2\pi$ or 256 pixels, as the grid resolution $N$ is $2^8$ pixels. Both the cases of co-axial jet are studied at Reynolds number corresponding to Re: 10,000. The inverse momentum thickness or the  steepness parameter $B = 10.5$, the jet diameter $r_o$ is kept constant at $\frac{L}{20}$.  The $U$,$V$ and $PS$ for a co-axial jet are defined as follows:

\begin{verbatim}
Umax = 0.1; Vmax=Umax/30; ro = L/20;
y1= L/2 - CASE; y2= L/2 + CASE;
U  = Umax*0.5*(1-tanh(B*(abs(Y-y1)./ro - ro./abs(Y-y1)))); 
U  = U+Umax*0.5*(1-tanh(B*(abs(Y-y2)./ro - ro./abs(Y-y2)))); 
U = Umax*U/max(max(U));
V  = Vmax*(rand(N,N)-0.5);
PS = U/max(max(U));
\end{verbatim}

Figure~\ref{comples5} shows the mean profile for the two cases investigated in this report. The interactions between the jets in case 1 starts at step 40, whereas in case 2 it is at time step 100.

\noindent The objectives of the paper is as follows:

\begin{itemize}
   \item POD and DMD is implemented with Matlab and used to analyse LES of subsonic jet.
  \item The potential of POD and DMD in the ICE applications is pointed out.
  \item Results for POD and DMD are discussed and compared.
 \end{itemize} 
\noindent The organisation of this report is as follows:
\noindent In section 2 we study the computational methods for POD and DMD. Section 3 deals with the experiments and implemention details. Finally, conclusions are drawn and further research directions are discussed in section 4.

\section{Computational Methods}
In this section, the POD and DMD methods are discussed.

\subsection{POD (Method of Snapshots)}
POD was first introduced in the field of Computational Fluid Dynamics (CFD) by Lumley~\cite{Lumleyberkooz1993proper}. POD is known by various names like KLD, Principle Componant Analysis (PCA) and Singular Value Decompostion (SVD).  This method aims at representing a 2D or 3D flow field in terms of orthogonal modes which correspond to  the energitic, coherent flow motions. Hence, POD may be used to extract certain derterministic features from turbulent data. POD theory is best described in~\cite{gorder}. The present day analysis uses method of snapshots introduced by Sirovich~\cite{sirovich1987turbulence}. Here LES data at a particular interval of time, interpolated on a uniform grid is considered to be one snapshot. This method was introduced to reduce the POD computations. To compute the POD using the  original method requires solving $n \times n$ eigenvalue problem. The main problem is the calculation of auto-covariance matrix $R$. Method of snapshots proposes that auto-covariance matrix can be approximated by a summation of 'snapshots'. Here $x$ is snapshot matrix.
\begin{equation}\label{eq:display7}
R_{ij} = \frac{1}{M} \sum_{n=1}^{M} x_i^nx_j^n 
\end{equation}
The snapshots are assumed to be distanced of spatial distance greater than the correlation distance. Nowadays this method has become extremely popular~\cite{duwig2010extended}~\cite{gorder}~\cite{meyer2007}. Its use in certain flows are questionable however, due to its assumption of the snapshots being uncorrelated. This discussion is briefly explained in ~\cite{gorder}.

\subsection{Dynamic Mode Decomposition (DMD)}
The mathematics underlying the extraction of dynamic information from time-resolved snapshots of
LES data is closely related to the idea underlying the Arnoldi algorithm.
\noindent Let $v_j$ denote each instantaneous flow field. A sequence of N is writen as :
\begin{equation}
V_1^N = [v_1 v_2 v_3 v_4 \cdots v_{N}]
\end{equation}
A linear mapping from one snapshot to another is assummed.
\begin{equation}
V_1^N = [v_1 Av_1 A^2v_1 A^3v_1 \cdots A^{N-1}v_{1}]
\end{equation}
This can further more assumed  to be a constant, so that the mapping A can be represented as: 
\begin{equation}
v_{j+1} = Av_j
\end{equation}
By the linear combination of available data fields, we have a standard Arlondi iteration problem ~\cite{schmid2010dynamic}.
\begin{equation}
Av_i^{N-1} \approx v_i^{N-1}S
\end{equation}
where S is a companion matrix that simply shifts the snapshots 1 through $N-1$ and approximates the last snapshot $N$ by a linear combination of previous $N-1$ snapshots. Hence this procedure will result in the low dimensional system matrix S. We solve the S matrix problem using eigenvalue analysis and obtain the eigenvalues. It is known that eigenvalues of S, approximate some of the eigenvalues of the full system A. The associated eigenvectors of S provide the coefficients of the linear combination that is necessary to express the modal structure in the snapshot basis. S matrix is calculated as follows:
\begin{equation}
S = R^{-1}Q^*v_N
\end{equation}
where $Q^*$ is the complex conjugate transpose of $Q$ from the QR-decomposition of $V_1^{N-1}$.

\section{Experiments and Implementation}
Here we discuss the Experimental and implementation details of POD and DMD on 2D co-axial  jet data.
\subsection{Pre-Processing}
\begin{figure}[htp!]
\centering
\begin{tabular}{cc}
\includegraphics[scale=0.4]{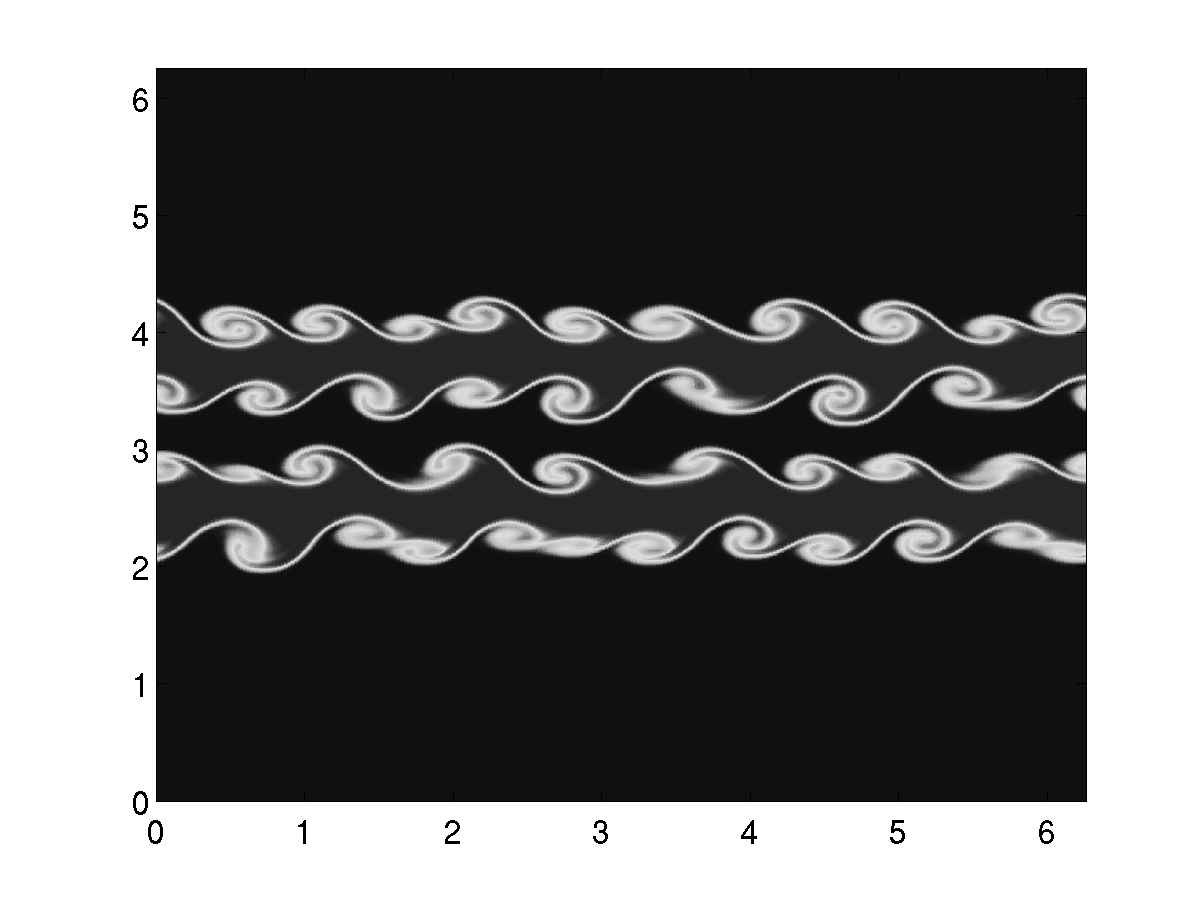} & \includegraphics[scale=0.4]{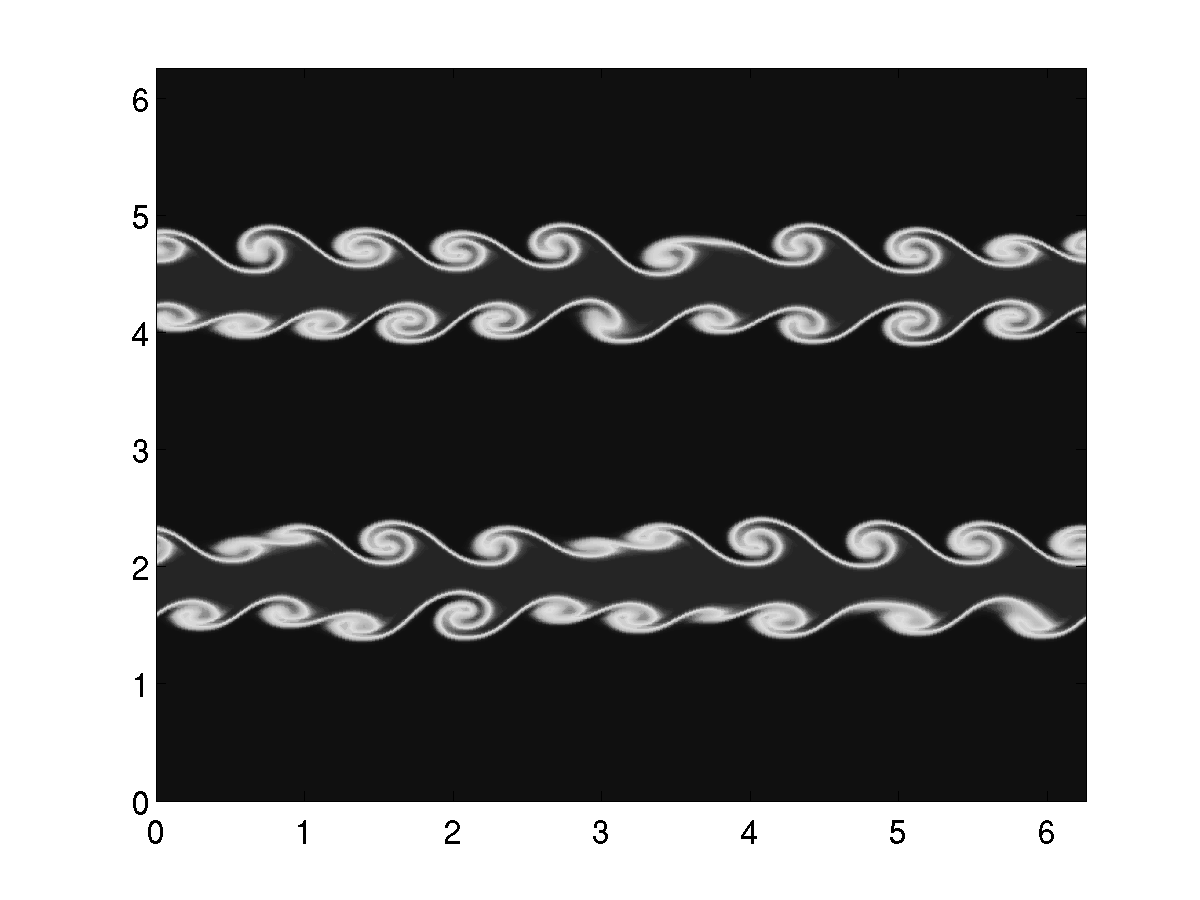}\\
\end{tabular}
\caption{case 1 (left) and case 2 (right). }
\label{comples1}
\end{figure}

Figure~\ref{comples1} shows 2D co-axial  jet for both the cases. A total of 30 snapshots were considered with a time interval of 5 steps.

Consider an example of a smoke in the wind flow, the concentration of the smoke $C$ at a particluar time t, is given by  $ \vec{C} = \vec{C}(P,t) $. As part of experiments, snapshot matrix $X$ is constructed at first. Let $P$ be the scalar concentration component:  
\begin{equation}
\label{eq:display7} 
C =   [P^1 P^2 P^3 \cdots P^N] = \begin{bmatrix}
P_{1}^1 & P_{1}^2 & ... & P_{1}^N \\
\vdots & \vdots  &  \vdots & \vdots  \\  
P_{M}^1 & P_{M}^2 & ... & P_{M}^N \\
\end{bmatrix}
\end{equation}

\subsection{Implementation of POD}
The fluctuating passive scalar matrix $U$ is calculated by substracting the mean from individual snapshots. Then the autocovariance matrix is computed as:
\begin{equation}\label{eq:display8}   
C = U^T U
\end{equation}

\noindent The eigenvalue problem for the matrix reads as follows:
\begin{equation}\label{eq:display9}   
C A^i = \lambda^i A^i
\end{equation}
The eigenvectors are arranged according to the decreasing order of eigenvalues reflecting the energies in different POD modes.

\begin{equation}\label{eq:display10}   
\lambda^1 > \lambda^2 > \lambda^3 > \lambda^4 > \cdots \lambda^N = 0.
\end{equation}
\noindent Using the ordered eigenvectors the POD modes are constructed.

\begin{equation}\label{eq:display11}   
\phi^i =  \frac{\sum_{n=1}^{N} A_n^i U^n}{\|\sum_{n=1}^{N} A_n^i U^n\|}, \hspace{6em} i = 1,2,\cdots,N.
\end{equation}

\subsection{Implementation of DMD}
\noindent Calculation of fluctuating passive scalar matrix is not needed. But the snapshot matrix is divided into two parts.
\begin{equation}\label{eq:display81}
V_1^{n-1} = [v_1 \;v_2 \; v_3  \; \cdots  \;v_{n-1}]
\end{equation}
\begin{equation}\label{eq:display18}
V_2^n = [v_2  \;v_3 \; v_4  \; \cdots \; v_{n}]
\end{equation}
QR decomposition in economy mode is  performed as :
\begin{equation}\label{eq:display19}
[Q,R] = qr(V_1^{n-1},0). 
\end{equation}
Companion matrix S is calculated as :
\begin{equation}\label{eq:display111}
S = R^{-1}Q^*V_2^n.
\end{equation}
The eigenvalue analysis is computed on S matrix
\begin{equation}\label{eq:display11}
[X,D] = eig(S)
\end{equation}
The dynamic mode spectrum is computed as :
\begin{equation}
\lambda_j = log(D_{jj})/\delta t.
\end{equation}
$\delta t$, is the time interval between the snapshots.
The Dynamic modes can be computed as follows:
\begin{equation}
DM_j = V_i^{N-1}X(:,j)
\end{equation}
where X is the original snapshot matrix. Dynamic mode decomposition contains not only information about coherent structures, but also about their temporal evolution. Since at no stage of the algorithm the system matrix A is needed, various extensions and attractive features of the algorithm should be noted. No specific spatial arrangement of the sampled data is
assumed, and the processing of subdomain data, i.e., data extracted only in a small region of the complete flow domain, as well as the processing of unstructured data is possible. This feature allows to probe a subregion of the flow as to the presence of a specific mechanism.

\section{Results and Discussion}
The POD modes are the optimal decomposition for the flow and capture large scale structures thereby providing information on the large scale behaviour. Total kinetic energy is contained within the first few POD modes. Generally 95\% of the total intensity fluctuations is used and the average flow field is described in the first $n$ POD modes. Dynamic modes represent the perturbation dynamics and capture the characteristic pattern located near the shear layer. DMD spectrum quantatively describes the jet behaviour. Higher the dynamic modes higher the is the frequency.

\begin{figure}[htp!]
\centering
\begin{tabular}{ccccc}
\includegraphics[scale=0.2]{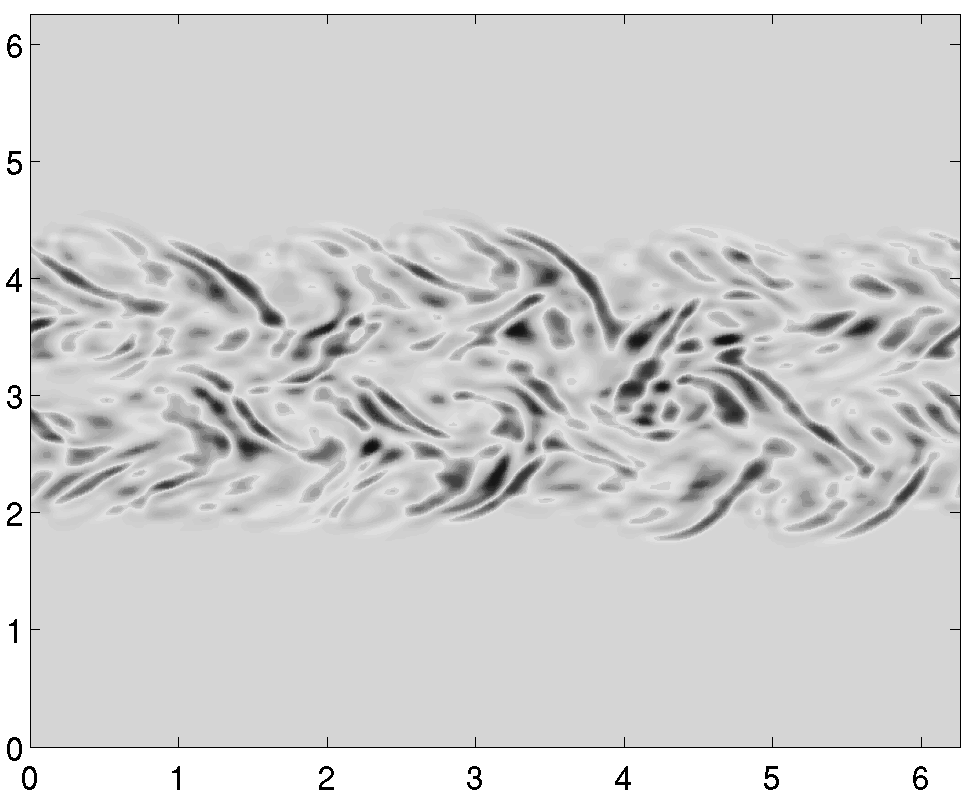} & \includegraphics[scale=0.2]{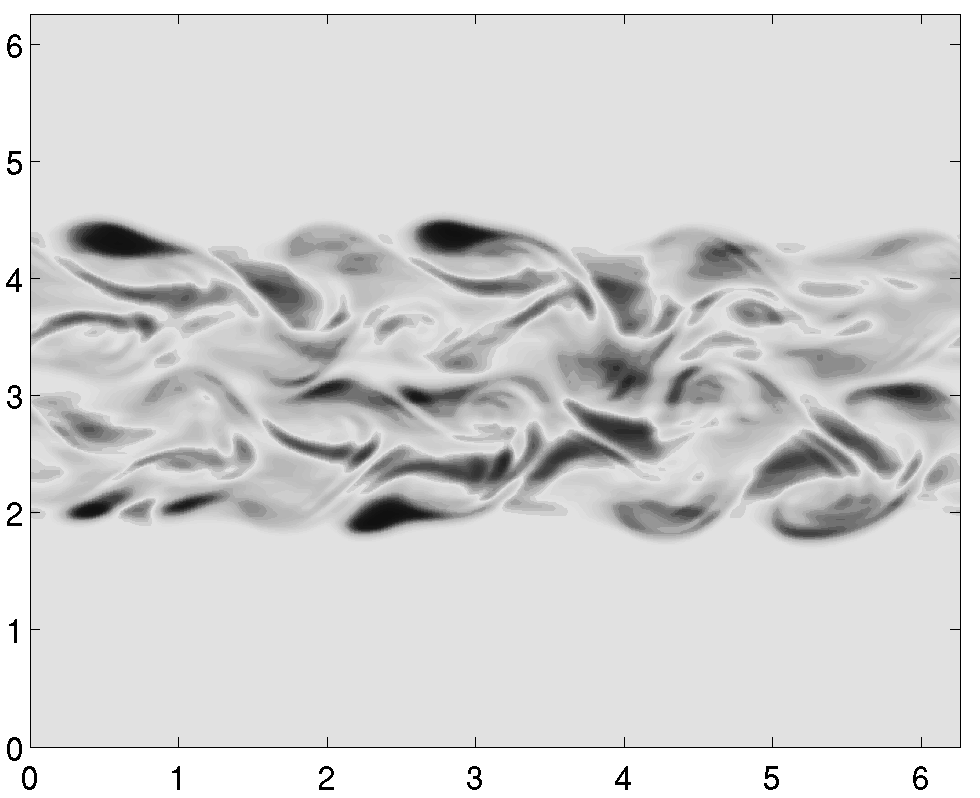} & \includegraphics[scale=0.2]{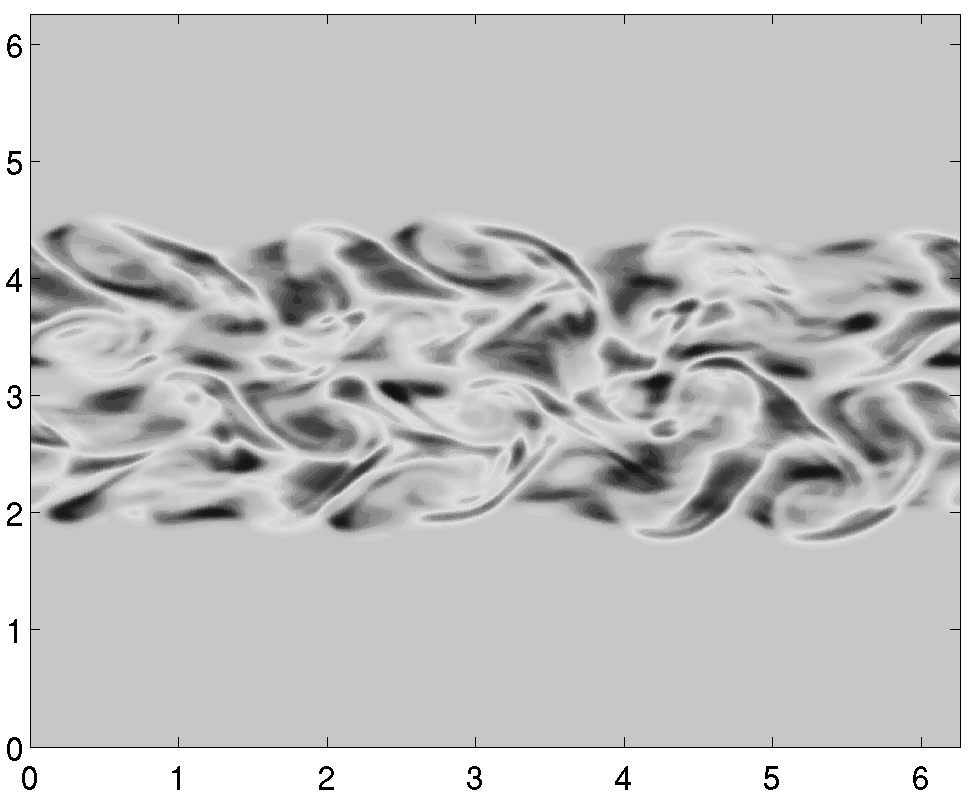} & \includegraphics[scale=0.2]{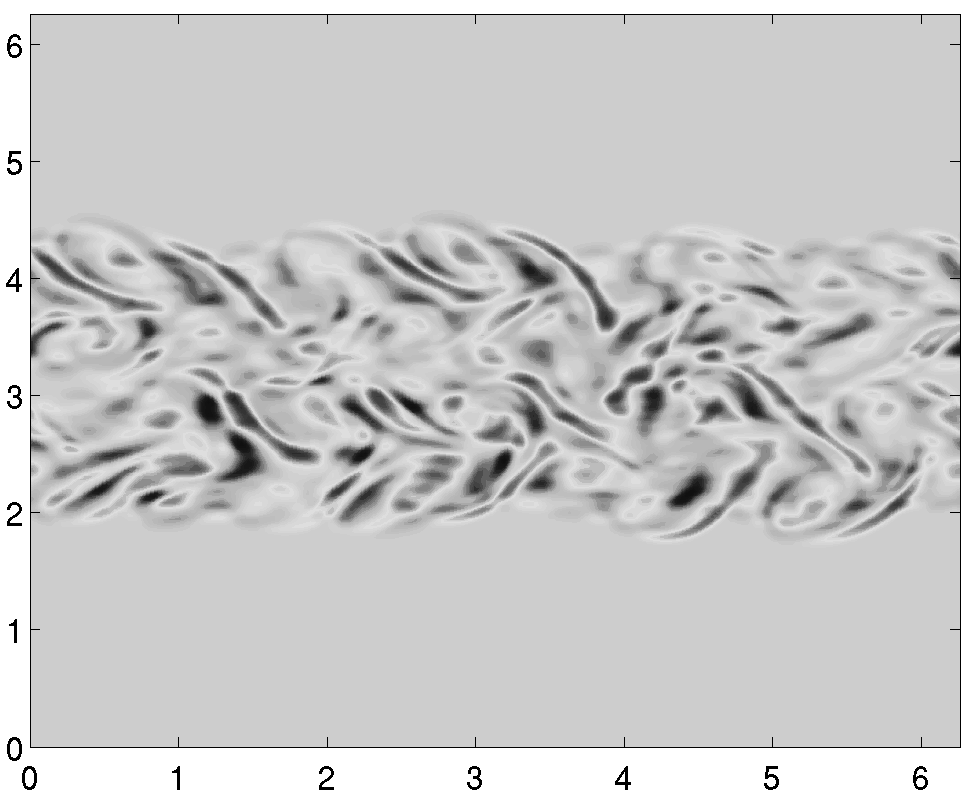} & \includegraphics[scale=0.2]{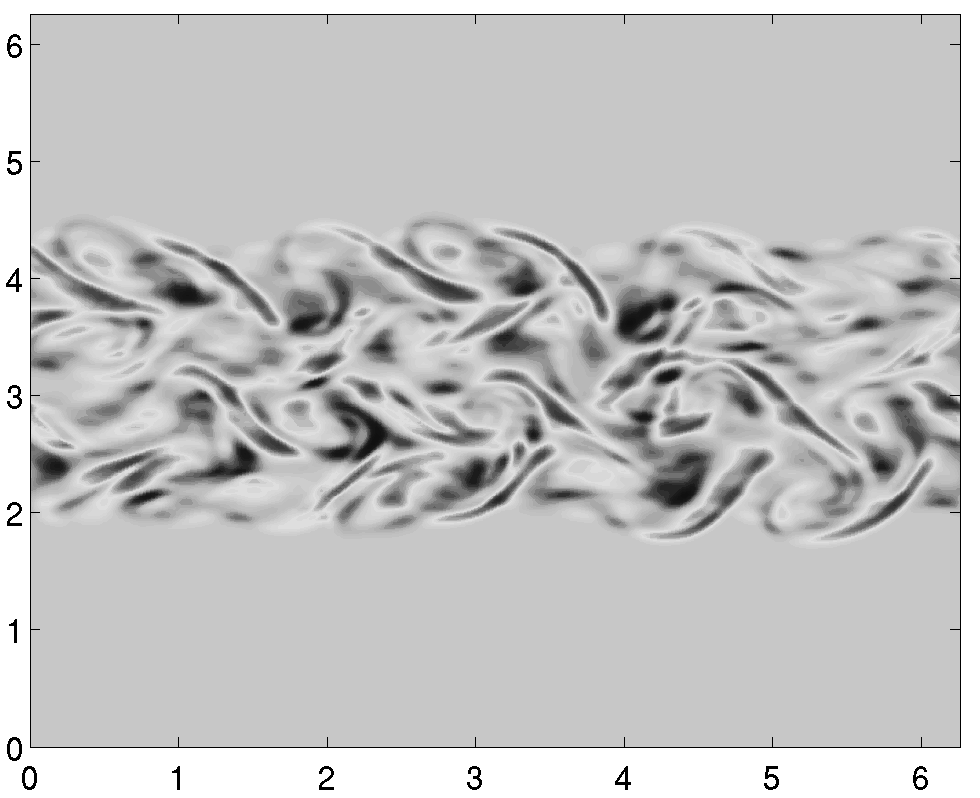}\\
\includegraphics[scale=0.2]{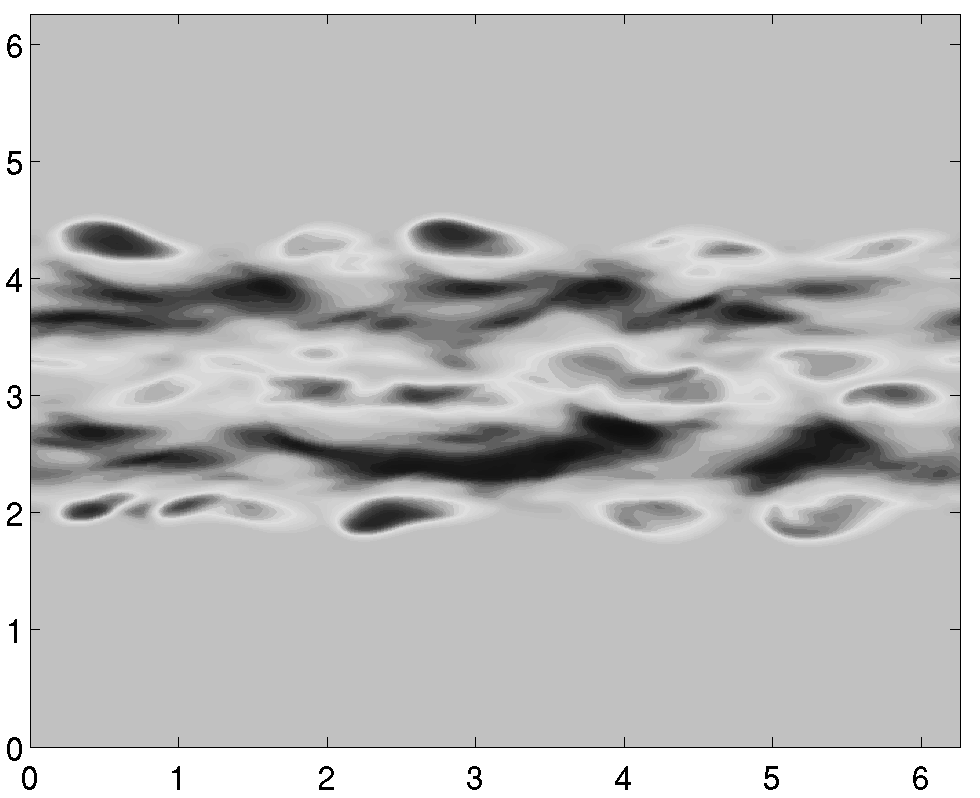} & \includegraphics[scale=0.2]{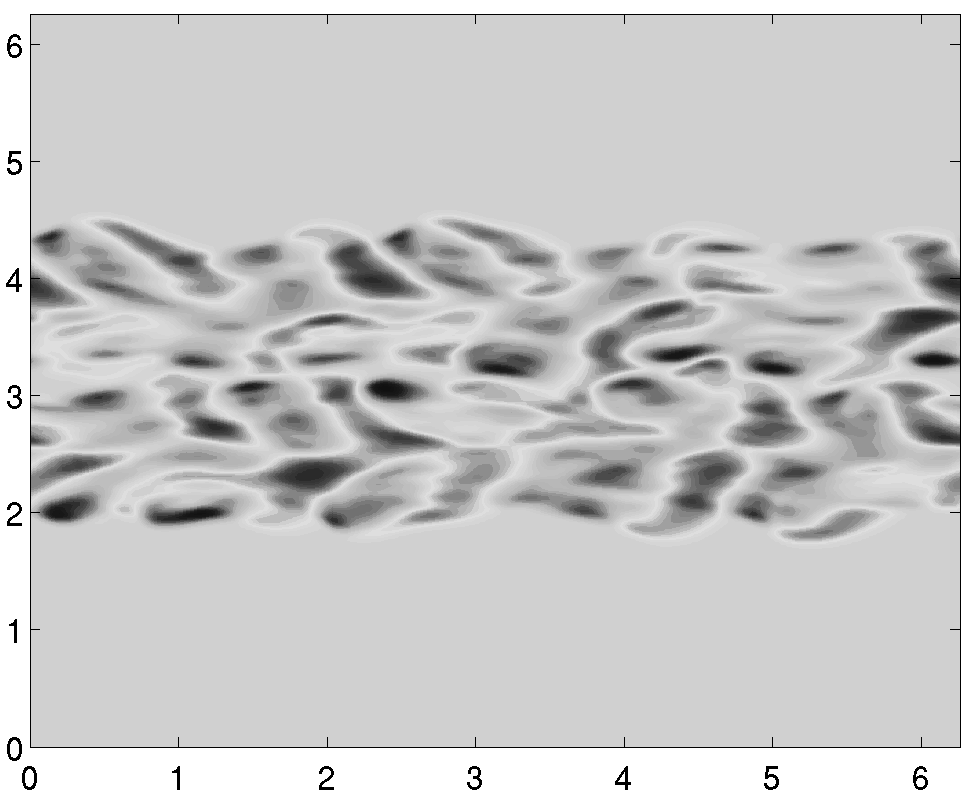} & \includegraphics[scale=0.2]{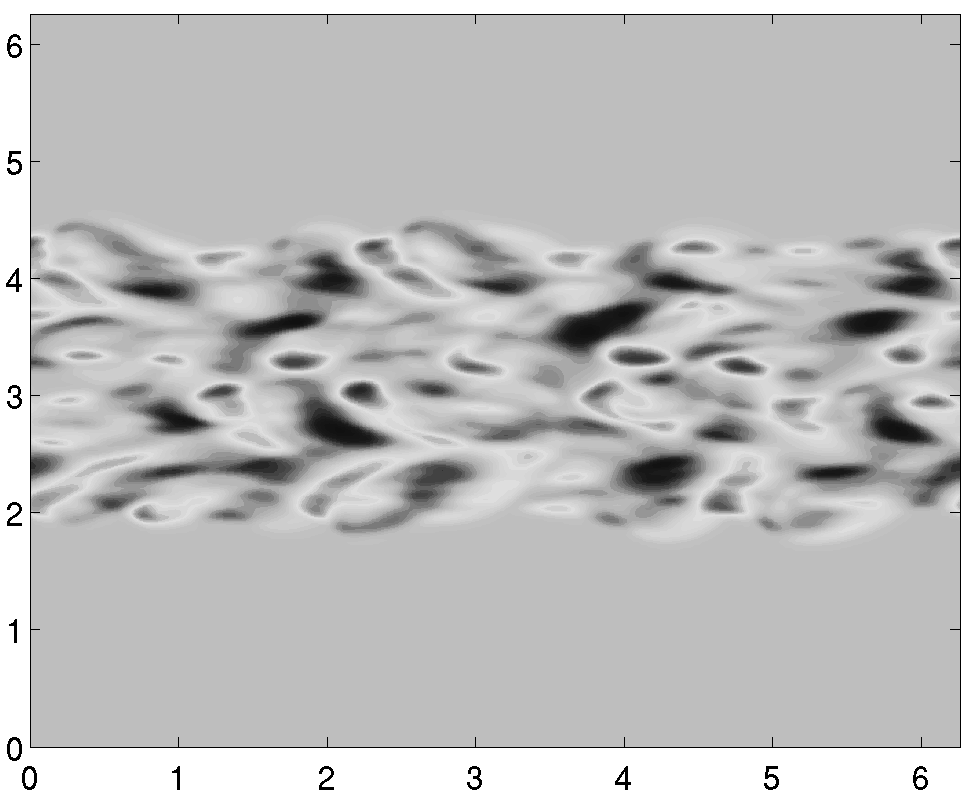} & \includegraphics[scale=0.2]{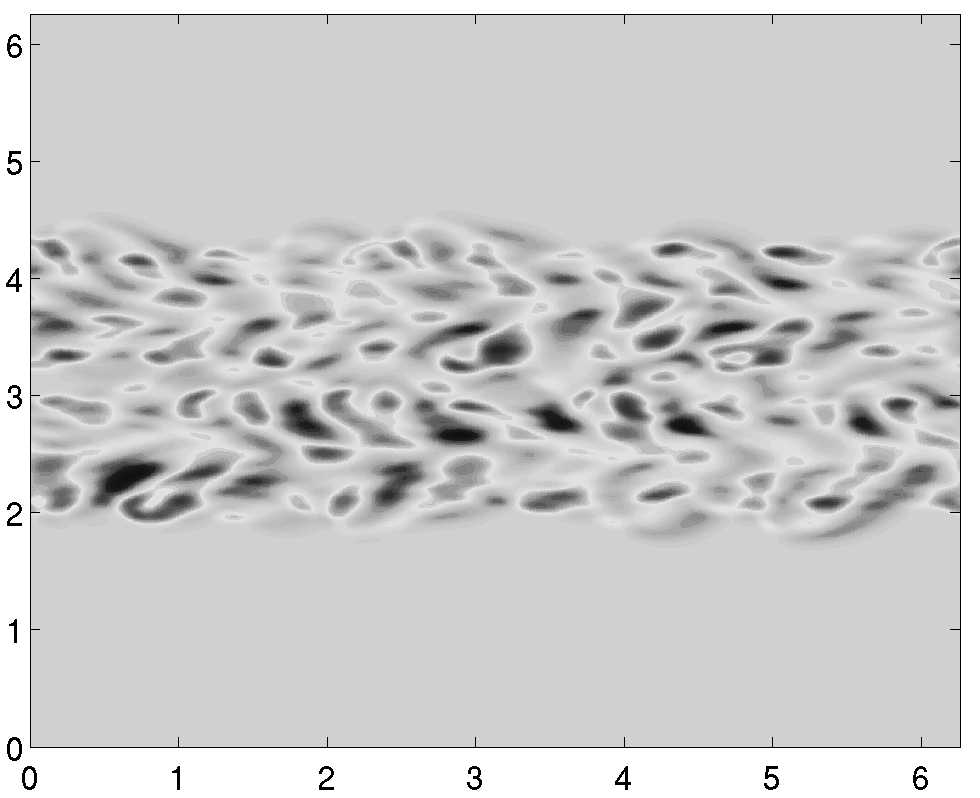} & \includegraphics[scale=0.2]{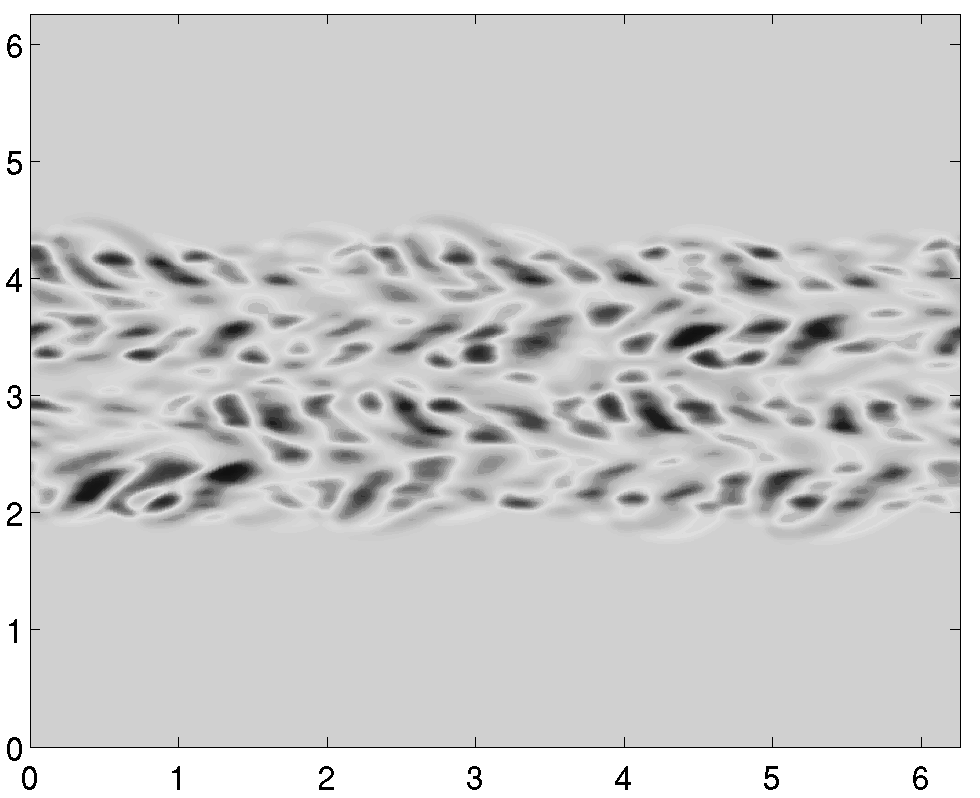}\\
\end{tabular}
\caption{Comparison of first 5 DMD modes (top) and POD modes (bottom) for case 1 passive scalar field.}
\label{comples2}
\end{figure}

\noindent From the Figure~\ref{comples2}, it is observed from POD modes, that there is a formation of symmetric structures of inverse Kelvin Helmholz instability due to interactions of both the jets. But from Dynamic modes, the inverse Kelvin Helmholz instability is clearly visible. The second dynamic mode shows a characteristic pattern located near the shear layer, which represents the roll-up of the symmetric vortex sheet in Figure~\ref{comples2}. The first and third mode depict small scale structures near the nozzle exit region.

\begin{figure}[htp!]
\centering
\begin{tabular}{ccccc}
\includegraphics[scale=0.2]{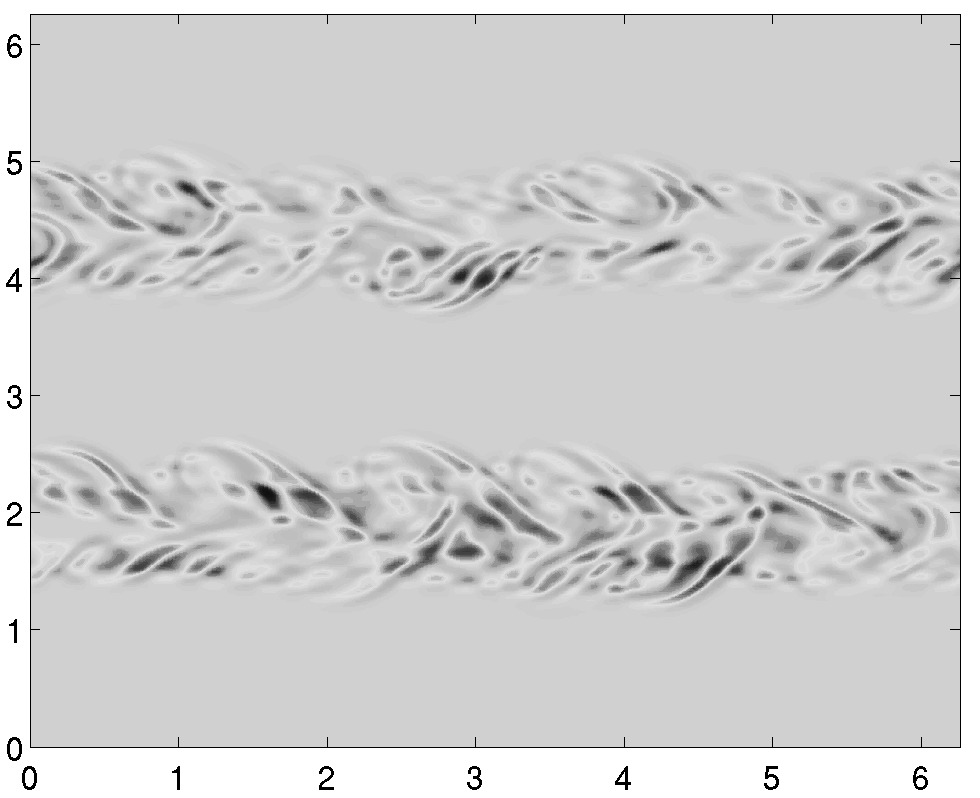} & \includegraphics[scale=0.2]{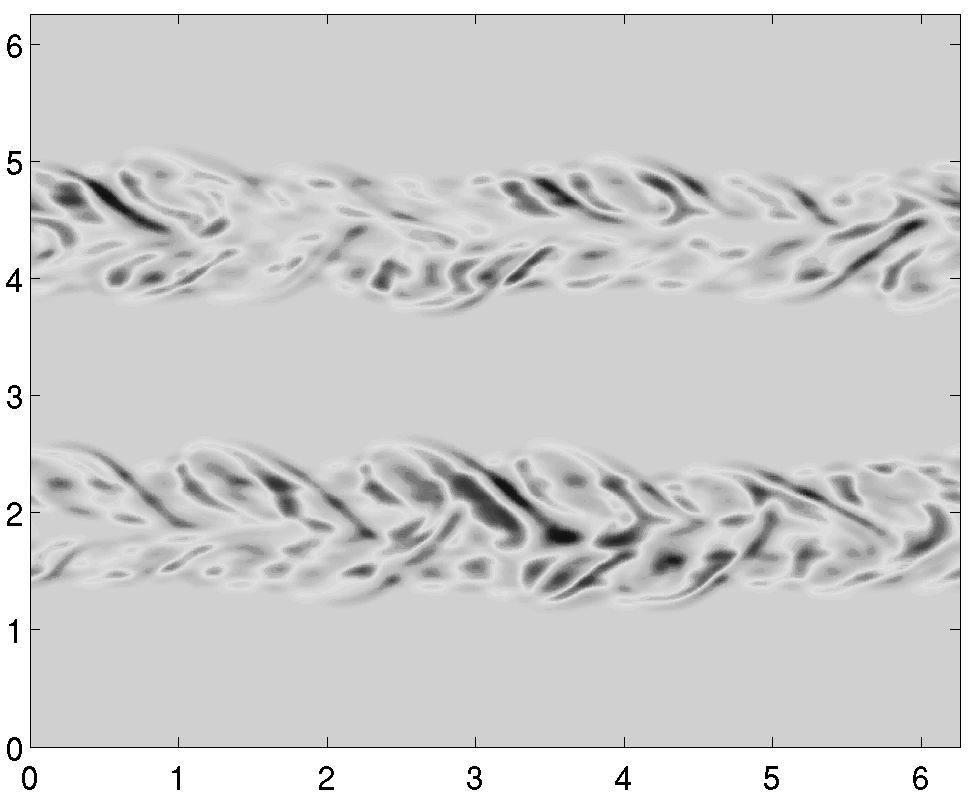} & \includegraphics[scale=0.2]{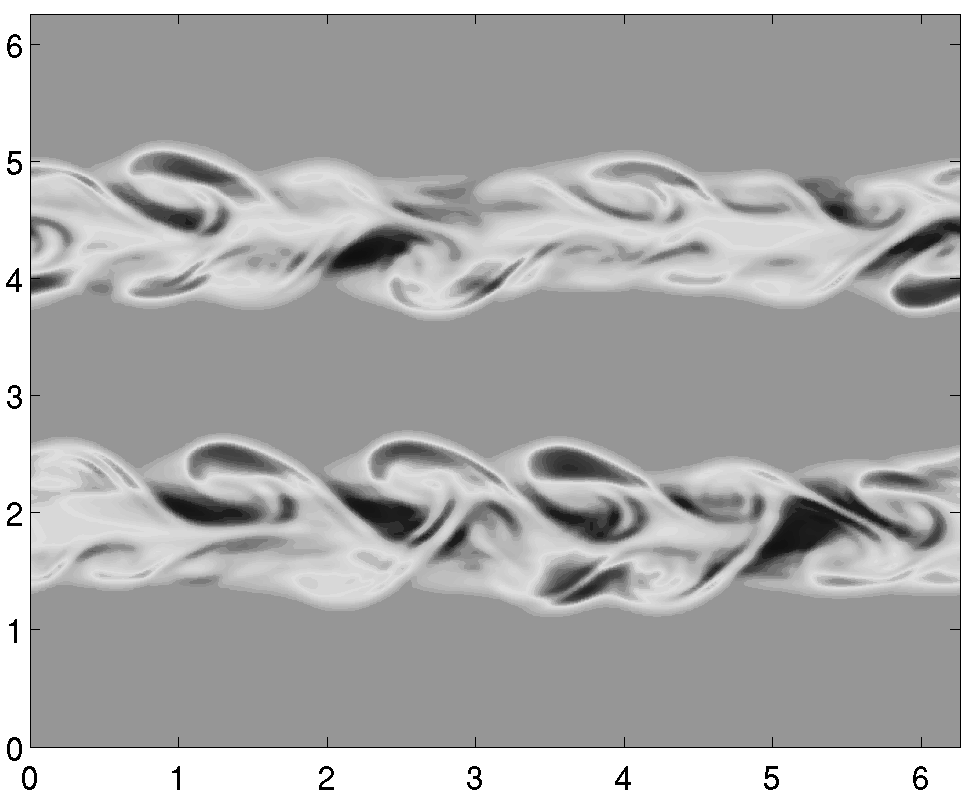} & \includegraphics[scale=0.2]{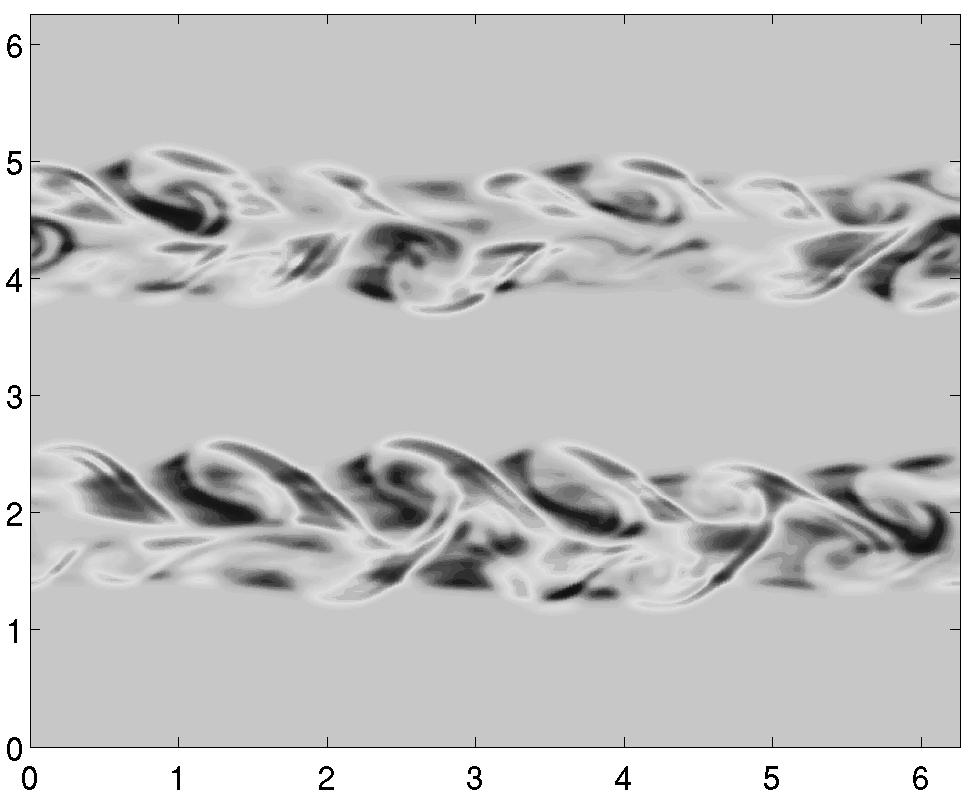} & \includegraphics[scale=0.2]{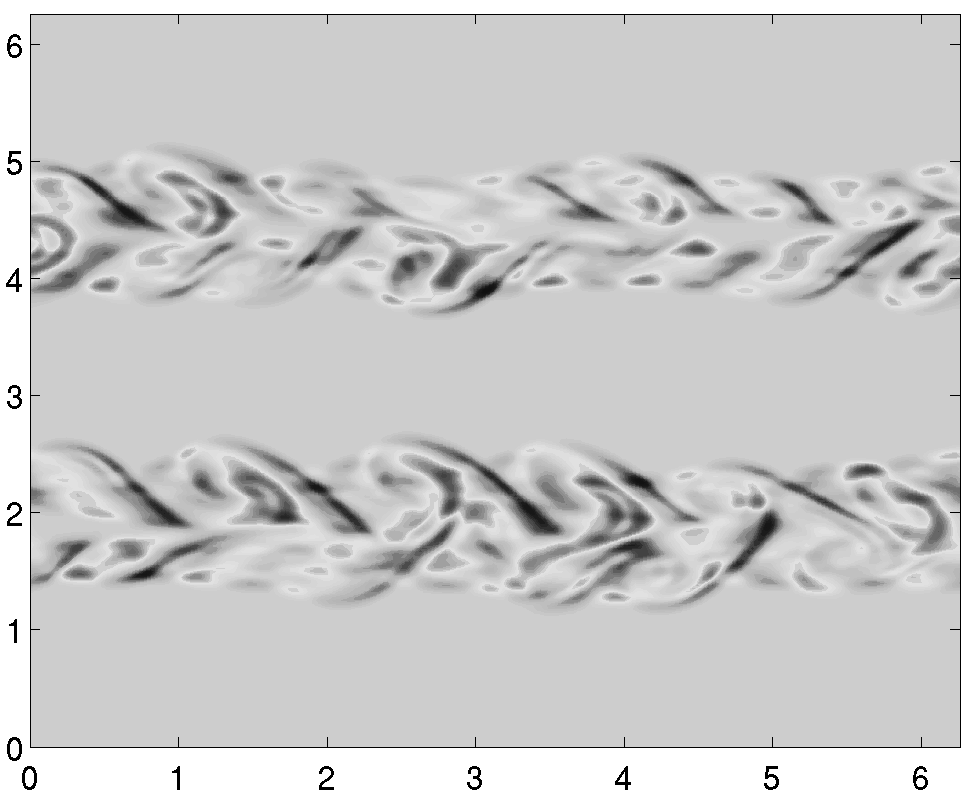}\\
\includegraphics[scale=0.2]{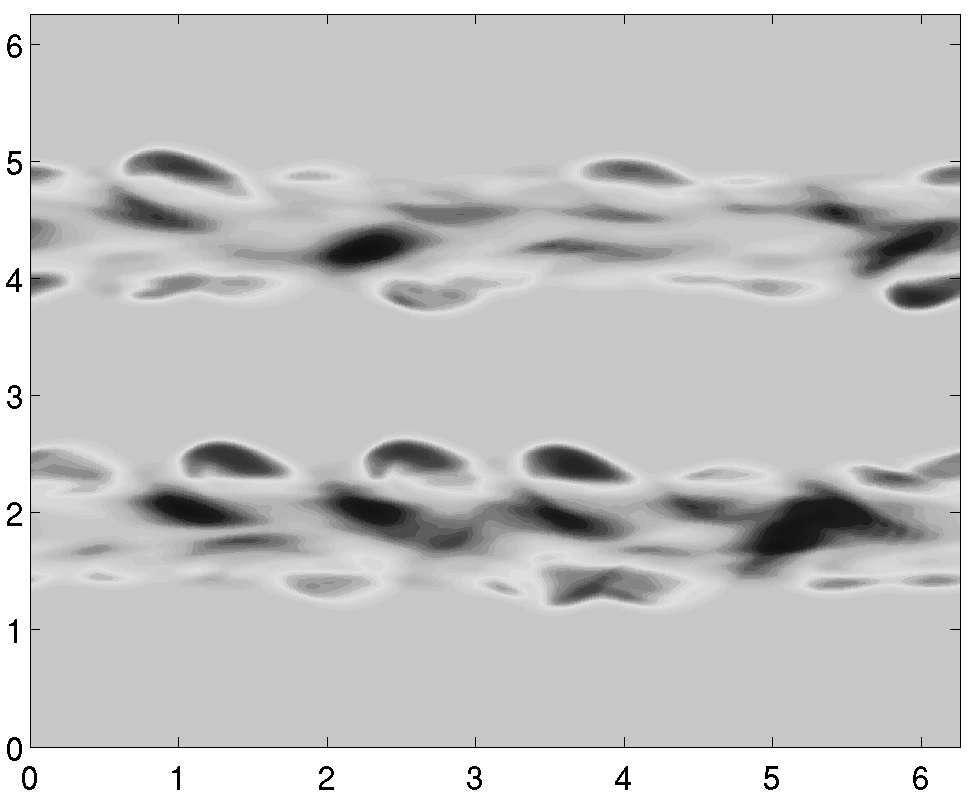} & \includegraphics[scale=0.2]{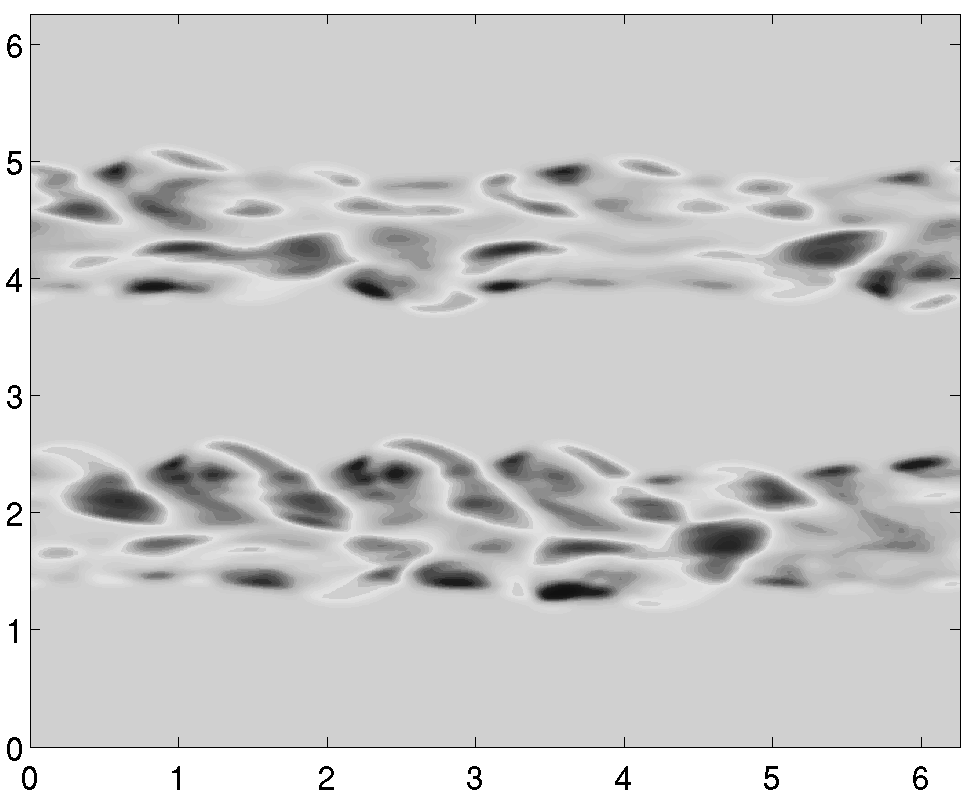} & \includegraphics[scale=0.2]{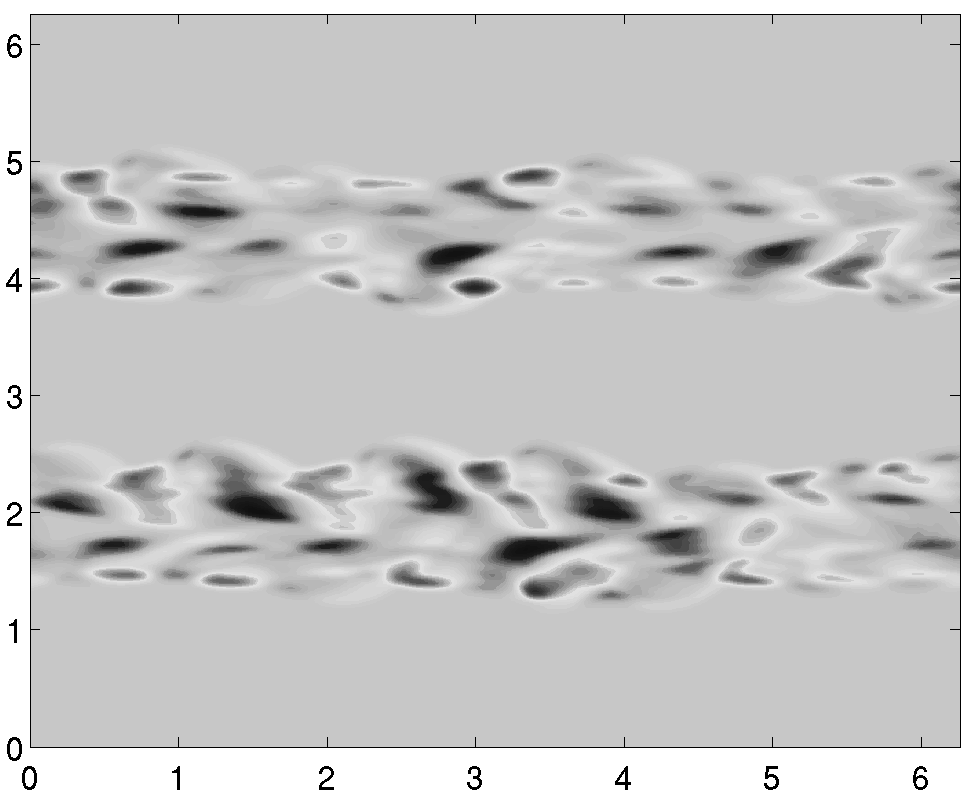} & \includegraphics[scale=0.2]{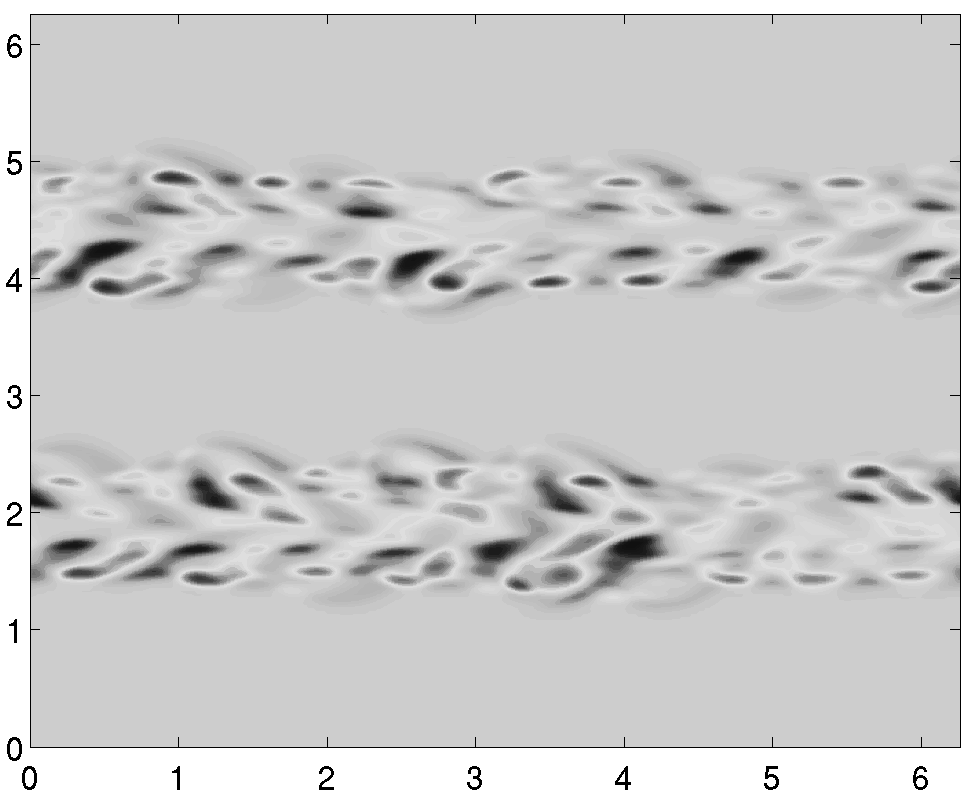} & \includegraphics[scale=0.2]{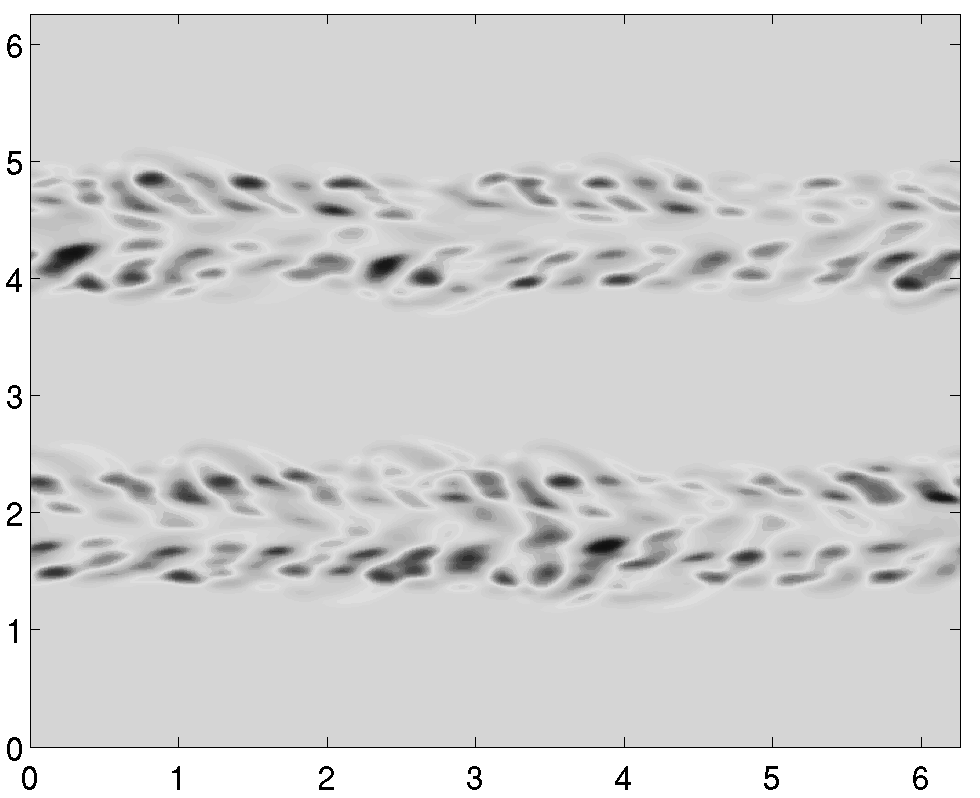}\\
\end{tabular}
\caption{Comparison of first 5 DMD modes (top) and POD modes (bottom) for case 2 passive scalar field.}
\label{comples3}
\end{figure}

\begin{figure}[htp!]
\centering
\begin{tabular}{cc}
\includegraphics[scale=0.5]{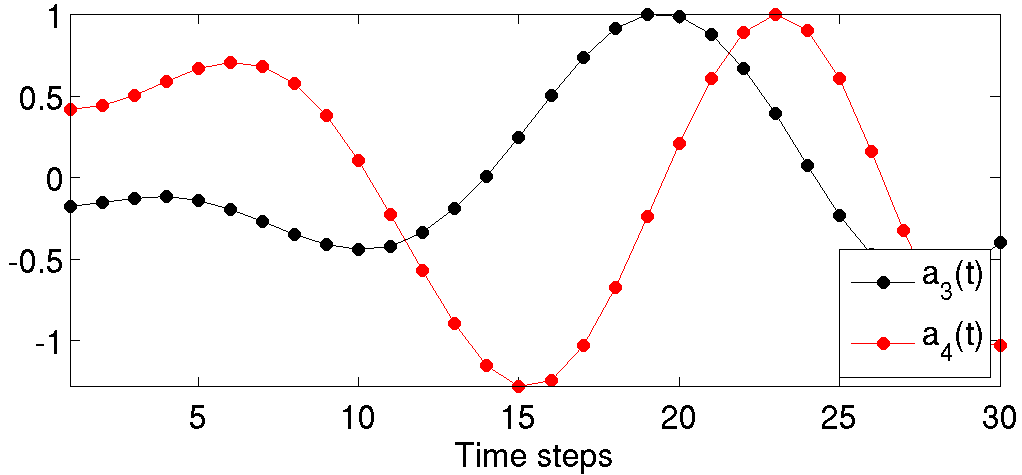} & \includegraphics[scale=0.5]{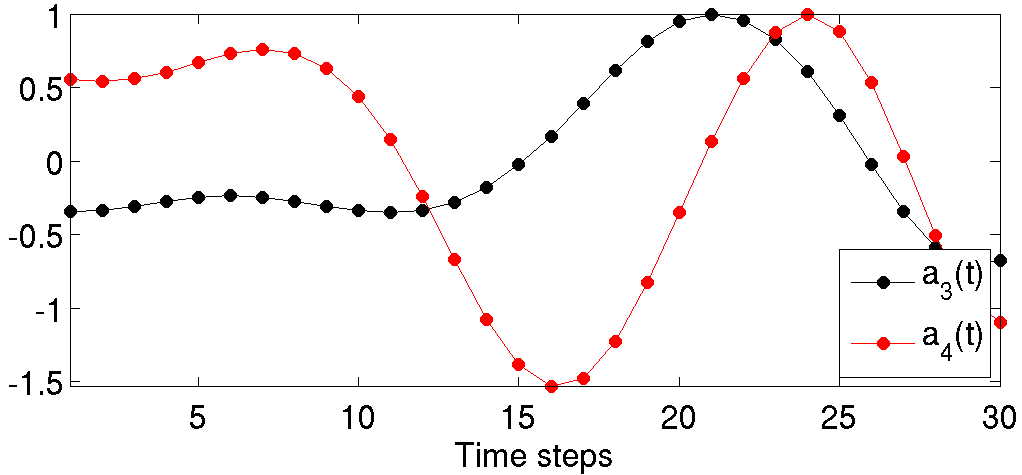}\\
\end{tabular}
\caption{Time coefficients of case 1(left) and case 2( right) passive scalar fields}
\label{comples4}
\end{figure}

\noindent From the Figure~\ref{comples3}, it is observed that there is the formation of the Kelvin Helmoltz instability in both the jets. From Figure~\ref{comples4}, the time coefficients of POD modes 3,4 show a phase difference of 5 time steps. In the other words, the POD modes 3,4 are strongly correlated to each other, This phenomena would further give rise to study helical and columnar features of instabilities.

\noindent POD and DMD methods can be utilised for analysing the LES data. The power of PODs fast convergence allows for the large scale structures to be isolated from the small scale structures in the turbulence. This would help in analysing the flow field in different ways. From the Figure~\ref{comples2} and Figure~\ref{comples3}, POD and DMD methods are shown as a powerful numerical tools, for use in jet flows. The ability to maximize the intensity fluctuations of the flow with a miminal number of modes shows PODs strength in the analysis of the coherent flow structures and reduced order modelling. The DMD method has clear advantages over POD, as it strives for a representation of the dominant flow features with in a temporal orthogonal framework,  while POD is based on a spatial orthogonal framework.  

\begin{figure}[htp!]
\centering
\begin{tabular}{c}
\includegraphics[scale=0.3]{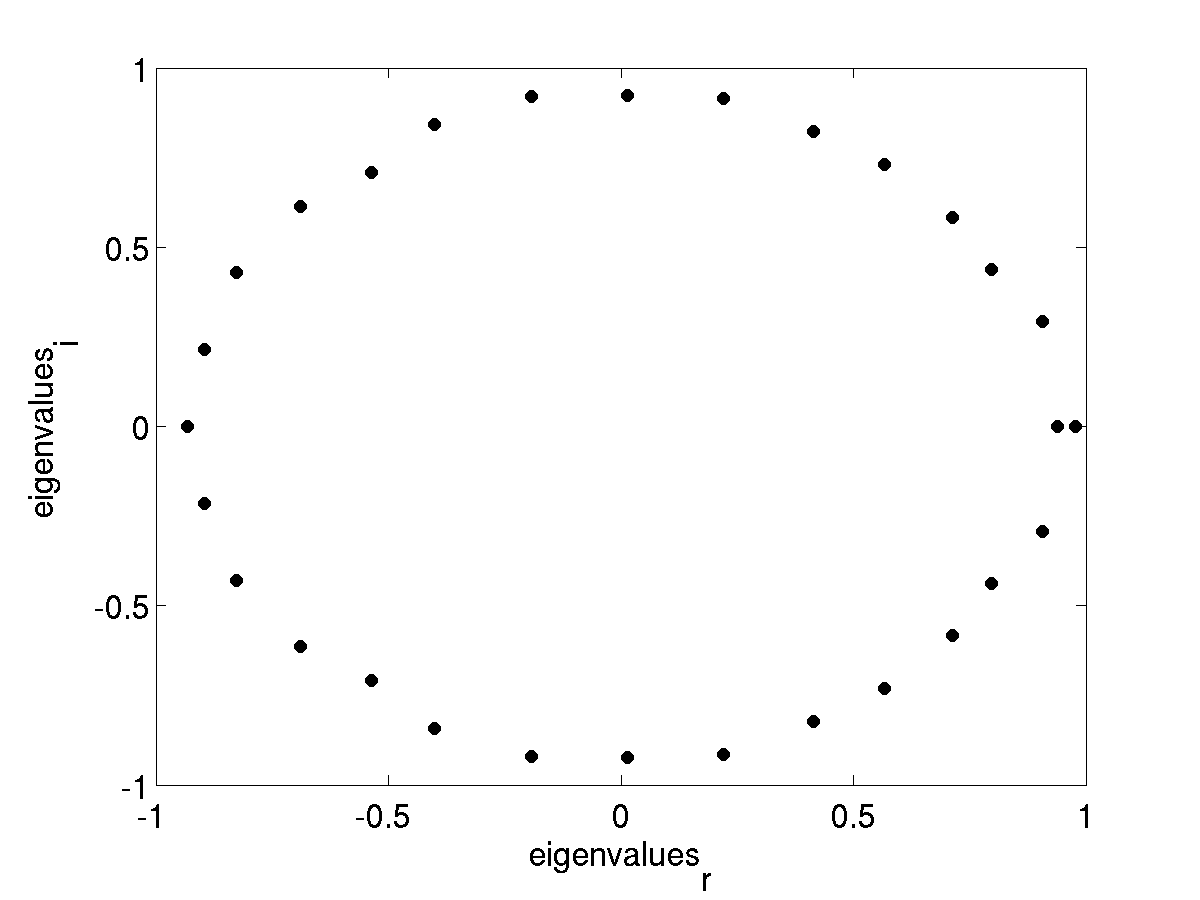}
\end{tabular}
\caption{Eigenvalues of DMD for case 1 passive scalar field.}
\label{comples6}
\end{figure}

\noindent From the Figure~\ref{comples6} the eigenvalues of S represent the mapping between subsequent snapshots: unstable eigenvalues are
given by a modulus greater than one  (i.e. $ |\lambda| > 1$, are located outside the unit disk); stable eigenvalues have
a modulus equal to one (i.e. $|\lambda| = 1$, can be found on the unit disk); unstable eigenvalues have
a modulus less than one (i.e. $|\lambda| < 1$, can be found inside the unit disk);. For applications in fluid dynamics,
it is common to transform the eigenvalues of S using a logarithmic mapping, after which the unstable (stable) eigenvalues have a positive (negative) real part. The procedural steps for computing the dynamic mode decomposition are given in section 3.

\section{Conclusions}
\noindent POD and DMD methods can be utilised for analysing the LES data. The power of PODs fast convergence allows for the large scale structures to be isolated from the small scale structures in the turbulence. The DMD method has clear advantages over POD, as it strives for a representation of the dominant flow features with in a temporal orthogonal framework,  while POD is based on a spatial orthogonal framework.  This paper summerizes that POD and DMD methods will provide the experimentalist with  solid tools, in quantifying important mechanisms in time resolved measurements of fluid dynamics. It is hoped that DMD and POD methods help in further understanding of fundamental fluid processes. In the future, effect of the time window between two frames of time resolved data on DMD method would be studied. Furthermore, the sensitivity analysis of the POD modes would be studied to show that the POD modes are independent of the number of snapshots.

\section*{Acknowledgement}

I thank Ville Vuorinen, D.Sc.(Tech.) for motivating me and helping me with solid ideas on how to proceed with this course work on 'Direct Numerical Simulations' 2012 at Aalto University, Finland. It would have been very difficult for me to understand the basics of the course work with out Ville's guidence.



%
\bibliographystyle{IEEEtran}
\bibliography{annot}

\end{document}